\documentclass[12pt,letterpaper]{article}

\usepackage{amsmath}
\usepackage{amsfonts}
\usepackage{amssymb}
\usepackage{graphicx}
\begin{document}


\begin{flushright} {\footnotesize MIT-CTP-3572}  \end{flushright}
\vspace{5mm} \vspace{0.5cm}
\begin{center}

\def\thefootnote{\fnsymbol{footnote}}

{\Large \bf How heavy can the Fermions in Split Susy be?\\ A study
on Gravitino and Extradimensional LSP.\\[1cm]}
{\large Leonardo Senatore\footnote{This work is supported in part
by funds provided by the U.S. Department of Energy (D.O.E) under
cooperative research agreement DF-FC02-94ER40818}}
 \vspace{.8cm}

{\small
\textit{Center for Theoretical Physics, \\
Massachusetts Institute of Technology, Cambridge, MA 02139, USA}}
\end{center}
\vspace{.2cm}
 \vspace{.8cm}

\hrule \vspace{0.3cm}
{\small  \noindent \textbf{Abstract} \\[0.3cm]
In recently introduced Split Susy theories, in which the scale of
Susy breaking is very high, the requirement that the relic
abundance of the Lightest SuperPartner (LSP) provides the Dark
Matter of the Universe leads to the prediction of fermionic
superpartners around the weak scale. This is no longer obviously
the case if the LSP is a hidden sector field, such as a Gravitino
or an other hidden sector fermion, so, it is interesting to study
this scenario. We consider the case in which the Next-Lightest
SuperPartner (NLSP) freezes out with its thermal relic abundance,
and then it decays to the LSP. We use the constraints from BBN and
CMB, together with the requirement of attaining Gauge Coupling
Unification and that the LSP abundance provides the Dark Matter of
the Universe, to infer the allowed superpartner spectrum. As very good
news for a possible detaction of Split Susy at LHC, we find that if the 
Gravitino is the LSP, than the only
allowed NLSP has to be very purely photino like. In this case, a photino
from 700 GeV to 5 TeV is allowed, which is difficult to test at LHC.  
We also study the
case where the LSP is given by a light fermion in the hidden
sector which is naturally present in Susy breaking in Extra
Dimensions. We find that, in this case, a generic NLSP is allowed
to be in the range 1-20 TeV, while a Bino NLSP can be as light as
tens of GeV.

\noindent

\vspace{0.5cm}  \hrule

\def\thefootnote{\arabic{footnote}}
\setcounter{footnote}{0}


\section{Introduction}
Two are the main reasons which lead to the introduction of Low
Energy Supersymmetry for the physics beyond the Standard Model: a
solution of the hierarchy problem, and gauge coupling unification.\\

The problem of the cosmological constant is usually neglected in
the general treatment of beyond the Standard Model physics,
justifying this with the assumption that its solution must come
from a quantum theory of gravity. However, recently
\cite{Arkani-Hamed:2004fb}, in the light of the landscape picture
developed by a new understanding of string theory, it has been
noted that, if the cosmological constant problem is solved just by
a choice of a particular vacua with the right amount of
cosmological constant, the statistical weight of such a fine
tuning may dominate the fine tuning necessary to keep the Higgs
light. Therefore, it is in this sense reasonable to expect that
the vacuum which solves the cosmological constant
problem solves also the hierarchy problem.\\

As a consequence of this, the necessity of having Susy at low
energy disappears, and Susy can be broken at much higher scales
($10^6-10^9$ GeV).\\

However, there is another important prediction of Low Energy Susy
which we do not want to give up, and this is gauge coupling
unification. Nevertheless, gauge coupling unification with the
same precision as with the usual Minimal Supersymmetric Standard
Model (MSSM) can be achieved also in the case in which Susy is
broken at high scales. An example of this is the theories called
Split Susy \cite{Arkani-Hamed:2004fb,Giudice:2004tc} where there
is an hierarchy between the scalar supersymmetric partners of
Standard Model (SM) particles (squarks, sleptons, and so on) and
the fermionic superpartners of SM particles (Gaugino, Higgsino),
according to which, the scalars can be very heavy at an
intermediate scale of the order of $10^9$ GeV, while the fermions
can be around the weak scale. The existence for this hierarchy can
be justified by requiring that the chiral symmetry protects the
mass of the
fermions partners.\\

While the chiral symmetry justifies the existence of light
fermions, it can not fix the mass of the fermionic partners
precisely at the weak scale. As a consequence, this theory tends
to make improbable the possibility of finding Susy at LHC, because
in principle there could be no particles at precisely 1 TeV. In
this paper, for Split Susy, we do a study at one-loop level of the
range of masses allowed by gauge coupling unification, finding
that these can vary in a range that approximately goes up to 20
TeV. A possible way out from this depressing scenario comes from
realizing that cosmological observations indicate the existence of
Dark Matter (DM) in the universe. The standard paradigm is that
the Dark Matter should be constituted by stable weakly interacting
particles which are thermal relics from the initial times of the
universe. The Lightest Supersymmetric Partner (LSP) in the case of
conserved R-parity is stable, and, if it is weakly interacting,
such as the Neutralino, it provides a perfect candidate for the
DM. In particular, an actual calculation shows that in order for
the LSP to provide all the DM of the universe, its mass should be
very close to the TeV scale. This is the very good news for LHC we
were looking for. Just to stress this result, it is the
requirement the the DM is given by weakly interacting LSP that
forces the fermions in Split Susy to be close to the weak scale,
and accessible at
LHC.\\

In three recent papers \cite{Giudice:2004tc,Pierce:2004mk,
Masiero:2004ft}, the
predictions for DM in Split Susy were investigated, and revealed
some regions in which the Neutralino can be as light as $\sim 200$
GeV (Bino-Higgsino), and some others instead where it is around a
1 TeV (Pure Higgsino) or even 2 TeV (Pure Wino). As we had
anticipated, all these scales are very close to one TeV, even
though only the Bino-Higgsino region is
very good for detection at LHC.\\

Since the Dark Matter Observation is really the constraint that
tells us if this kind of theories will be observable or not at
LHC, it is worth to explore all the possibilities for DM in Split
Susy. In particular, a possible and well motivated case which had
been not considered in the literature, is the case in
which the LSP is a very weakly interacting fermion in a hidden sector.\\

In this paper, we will explore this possibility in the case in
which the LSP is either the Gravitino, or a light weakly
interacting fermion in the hidden sector which naturally appears
in Extra Dimensional Susy breaking models of Split Susy
\cite{Arkani-Hamed:2004fb,Luty:2002hj}.\\

We will find that, if the Gravitino is the LSP, than all possible candidates 
for the NLSP are excluded by the combination of imposing gauge
coupling unification and the constraint on hadronic decays coming 
from BBN. Just the requirement of having the Gravitino
to provide all the Dark Matter of the univese 
and to still have gauge coupling unification would have allowed weakly 
interacting fermionic superpartneres as heavy as 5 TeV, with very bad
consequences on the detactibility of Split Susy at LHC. This 
means that these constraints play a very big role. The only 
exception to this result occurs if the  NLSP is very photino like, 
avoiding in this way the stringent constraints on hadronic decays coming
from BBN. However, as we will see, already a small barionic decay branching
ratio of $10^{-3}$ is enough to rule out also this possibility.\\

For the Extradimensional LSP, we will instead find a wide range of 
possibilities, with NLSP allowed to span from 30 GeV to 20 TeV.\\
 
The paper is organized as follows. In section 2, we study the constraints
on the spectrum coming from the requirement of obtaining gauge coupling 
unification. In section 3, we briefly review the relic abundance of Dark 
Matter in the case the LSP is an hidden sector particle. In section 4, 
we discuss the cosmological constraints coming from BBN and CMB. In section 5, 
we show the results for Gravitino LSP. In section 6, we do the same for 
a dark sector LSP arising in extra dimensional implementation of Split Susy.
In section 7, we draw our conclusions.

\section{Gauge Coupling Unification}

Gauge coupling unification is a necessary requirement in Split
Susy theories. Here we investigate at one loop level how heavy can
be the fermionic supersymmetric partner for which gauge coupling
unification is allowed. We will consider the Bino, Wino, and
Higgsino as degenerate at a scale $M_2$, while we will put the
Gluinos at a different scale $M_3$.\\

Before actually beginning the computation, it is interesting to
make an observation about the lower bound on the mass of the
fermionic superpartners. Since the Bino is gauge singlet, it has
no effect on one-loop gauge coupling unification. In Split Susy,
with the scalar superpartners very heavy, the Bino is very weakly
interacting, its only relevant vertex being the one with the light
Higgs and the Higgsino. This means that, while for the other
supersymmetric partners LEP gives a lower bound of $\sim$ 50-100
GeV \cite{Eidelman:2004wy}, for the Bino in Split Susy there is
basically no
lower limit.\\

Going back to the computation of gauge coupling unification, we
perform the study at 1-loop level. The renormalization group
equations for the gauge couplings are given by:
\begin{equation}
\Lambda \frac{d g_i}{d\Lambda}=\frac{1}{(4\pi)^2}b_i(\Lambda)
g^3_i \label{rge}
\end{equation}
where $b_i(\Lambda)$ depends of the scale, keeping truck of the
different particle content of the theory according to the
different scales, and $i=1,2,3$ represent respectively
$\sqrt{5/3}g',g,g_s$. We introduce two different scales for the
Neutralinos, $M_2$, and
for the Gluinos $M_3$, and for us $M_3>M_2$.\\

In the effective theory below $M_2$, we have the SM, which
implies:
\begin{equation}
b^{SM}=(\frac{41}{10},-\frac{19}{7},-7)
\end{equation}
Between $M_2$ and $M_3$:
\begin{equation}
b^{split1}=(\frac{9}{2},-\frac{7}{6},-7)
\end{equation}
Between $M_3$ and $\tilde{m}$, which is the scale of the scalars:
\begin{equation}
b^{split2}=(\frac{9}{2},-\frac{7}{6},-5)
\end{equation}
and finally, above $\tilde{m}$ we have the SSM:
\begin{equation}
b^{ssm}=(\frac{33}{5},1,-3)
\end{equation}

The way we proceed is as follows: we compute the unification scale
$M_{GUT}$ and $\alpha_{GUT}$ as deduced by the unification of the
SU(2) and U(1) couplings. Starting from this, we deduce the value
of $\alpha_s$ at the weak scale $M_Z$, and we impose it to be
within the $2\sigma$ experimental result $\alpha_s(M_Z)=0.119\pm0.003$. We
use the experimental data:
$\sin^2(\theta_W(M_Z))=0.23150\pm0.00016$ and
$\alpha^{-1}(M_Z)=128.936\pm0.0049$\cite{Altarelli:2004fq}.\\

A further constraint comes from Proton decay $p\rightarrow
\pi^0e^+$, which has lifetime:
\begin{eqnarray}
&&\tau(p\rightarrow\pi_0e^+)=\frac{8f^2_\pi M^4_{GUT}}{\pi m_p
\alpha^2_{GUT}((1+D+F)A\alpha_N)^2}=\\ \nonumber
&&=\left(\frac{M_{GUT}}{10^{16} {\rm
GeV}}\right)^4\left(\frac{1/35}{\alpha_{GUT}}\right)^2\left(\frac{0.015 {\rm
GeV}^3}{\alpha_N}\right) 1.3\times 10^{35}{\rm yr}
\end{eqnarray}
where we have taken the chiral Lagrangian factor $(1+D+F)$ and the
operator renormalization $A$ to be $(1+D+F)A\simeq 20$. For the
Hadronic matrix element $\alpha_N$, we take the lattice result
\cite{Aoki:1999tw} $\alpha_N=0.015 {\rm GeV^3}$. From the
Super-Kamiokande limit \cite{Suzuki:2001rb},
$\tau(p\rightarrow\pi^0e^+)>5.3\times 10^{33}$yr, we get:
\begin{equation}
M_{GUT}>\left(\frac{\alpha_N}{0.015 {\rm
GeV}^3}\right)^{1/2}\left(\frac{\alpha_{GUT}}{1/35}\right)^{1/2} 4\times10^{15}
{\rm GeV}
\end{equation}

An important point regards the mass thresholds of the theory. In
fact, the spectrum of the theory will depend strongly on the
initial condition for the masses at the supersymmetric scale
$\tilde{m}$.  As we will see, in particular, the Gluino mass $M_3$
has a very important role for determining the allowed mass range
for the Next-Lightest Supersymmetric Particle (NLSP),
which is what we are trying to determine. In the
light of this, we will consider $M_2$ as a free parameter, with
the only constraint of being smaller than $\tilde{m}$. $M_3$ will
be then a function of $M_2$ and $\tilde{m}$, and its actual value
will depend on the kind of initial conditions we require. In order
to cover the larger fraction of parameter space as possible,
we will consider two distinct and
well motivated initial conditions. First, we will require gaugino
mass unification at $\tilde{m}$. This initial condition is the
best motivated in the approach of Split Susy, where
unification plays a fundamental role. Secondarily, we will require
anomaly mediated gauigino mass initial conditions at the scale
$\tilde{m}$. This second kind of initial conditions will give
results quite different from those of Gaugino mass unification,
and, even if in this case the Gravitino can not be the NLSP, the
field $\psi_X$, which will be a canditate LSP from extradimensions
that we will introduce in the next sections, could be still the
LSP.\\

\subsection{Gaugino Mass Unification}

Here we study the case in which we apply gaugino mass unification
at the scale $\tilde{m}$.\\

In \cite{Giudice:2004tc}, a 2-loop study of the renormalization
group equations for the Gaugino mass starting from this initial
condition was done, and it was found that, according to
$\tilde{m}$ and $M_2$, the ratio between $M_3$ and $M_2$ can vary in a
range $\sim 3-8$. We shall use their result for $M_3$, as the value of $M_3$
will have influence on the results, tending to increase the upper
limit on
the fermions' mass.\\

At one loop level, we can obtain analytical results. After
integration of eq.(\ref{rge}), we get the following expressions:
\begin{eqnarray}
M_{GUT}=&&\Big(e^{8\pi^2\frac{1}{g^2_1(M_Z)}-
\frac{1}{g^2_2(M_Z)}}M^{\left(b^{sm}_1-b^{sm}_2\right)}_Z
M^{\left((b^{split1}_1-b^{sm}_1)-(b^{split1}_2-b^{sm}_2)\right)}_2\\ \nonumber
&& M^{\left((b^{split2}_1-b^{split1}_1)-(b^{split2}_2-b^{split1}_2)\right)}_3
\tilde{m}^{\left((b^{ssm}_1-b^{split2}_1)-(b^{ssm}_2-b^{split2}_2)\right)}
\Big)^{\left(\frac{1}{b^{ssm}_1-b^{ssm}_2}\right)}
\end{eqnarray}
\begin{eqnarray}
\frac{1}{g^2_{GUT}}=&&\frac{1}{g^2_2(M_Z)}-\frac{1}{8\pi^2}\ln\big(M^{ -b^{sm}_2}_Z
M^{(-b^{split1}_2+b^{sm}_2)}_2\\ \nonumber
&&M^{(b^{split2}_2+b^{split1}_2)}_3
\tilde{m}^{(-b^{ssm}_2+b^{split2}_2)} M^{b^{ssm}_2}_{GUT}\big)
\end{eqnarray}

\begin{eqnarray}
\frac{1}{g^2_{s}(M_Z)}=&&\frac{1}{g^2(M_{GUT})}+\frac{1}{8\pi^2}\ln\Big
(M^{-b^{sm}_3}_Z
M^{(-b^{split1}_3+b^{sm}_3)}_2\\ \nonumber
&&M^{(-b^{split2}_3+b^{split1}_3)}_3
\tilde{m}^{(-b^{ssm}_3+b^{split2}_3)} M^{b^{ssm}_3}_{GUT}\big)
\end{eqnarray}

It turns out that two loops effect are important to determine the
predicted value of $\alpha_s(M_Z)$. Since our main purpose is to
have a rough idea of the maximum scale for the fermionic masses
allowed by Gauge Coupling Unification, we proceed in the following
way. In \cite{Giudice:2004tc}, 2-loop gauge coupling unification
was studied for $M_2=300$ GeV and 1 TeV. Since the main effect of
the 2-loop contribution is to raise the predicted value of
$\alpha_s(M_Z)$, we translate our predicted value of
$\alpha_s(M_Z)$ to match the result in \cite{Giudice:2004tc} for
the correspondent values of $M_2$. Having set in this way the
predicted scale for $\alpha_s(M_Z)$, we check what is the upper
limit on fermion masses in order to reach gauge coupling
unification. The amount of translation we have to do is: 0.008.\\

In fig.\ref{Gut_Unification}, we plot the prediction for
$\alpha_s(M_z)$ for $M_2=300$ GeV, 1 TeV, and 5 TeV. We see that
for 5 TeV, unification becomes impossible. And so, 5 TeV is the
upper limit on fermionic superpartner allowed from gauge coupling
unification. Note that the role of the small difference between
$M_3$ and $M_2$ is to raise this limit.\\

\begin{figure}[!h]
\begin{center}
\includegraphics[scale=0.85]{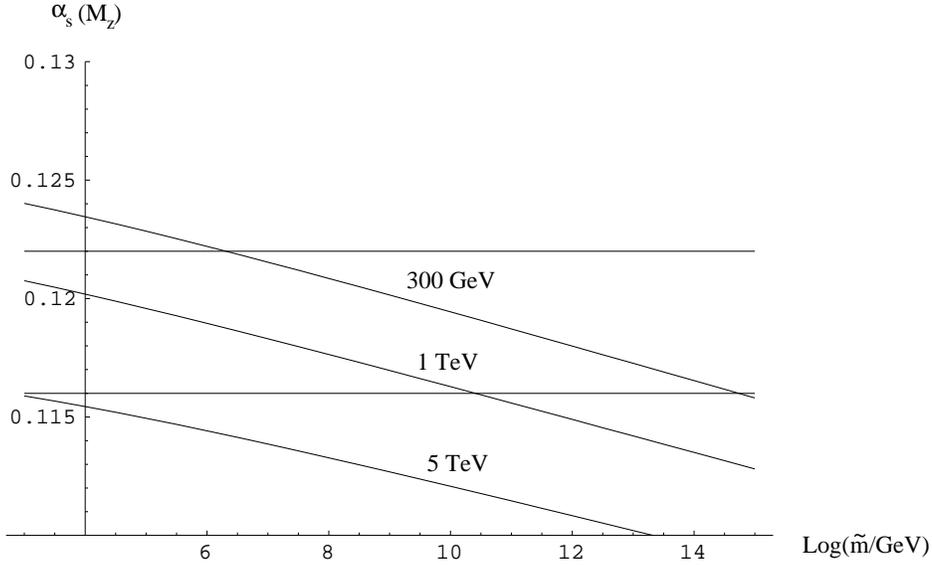}
\caption{In the case of gaugino mass unification at scale
$\tilde{m}$, we plot the unification prediction for
$\alpha_s(M_Z)$. The results for $M_2=300$ GeV, 1 TeV and 5 TeV
are shown. The horizontal lines represent the experimental bounds}
\label{Gut_Unification}
\end{center}
\end{figure}

In fig.\ref{alphaGut_Unification}, and
fig.\ref{Mgut_Unification},we plot the predictions for
$\alpha_{GUT}(M_{GUT})$ and for $M_{GUT}$, for the same range of
masses. We see that unification is reached in the perturbative
regime, with unification scale large enough to avoid proton decay
limits. Note, however, that for $M_2=5$ TeV, the limit is close to a
possible detection.\\

Finally, note that with this Gaugino mass initial conditions, the
Wino can not be the
NLSP if the Gravitino is the LSP, as shown in \cite{Giudice:2004tc}.\\

\begin{figure}[!h]
\begin{center}
\includegraphics{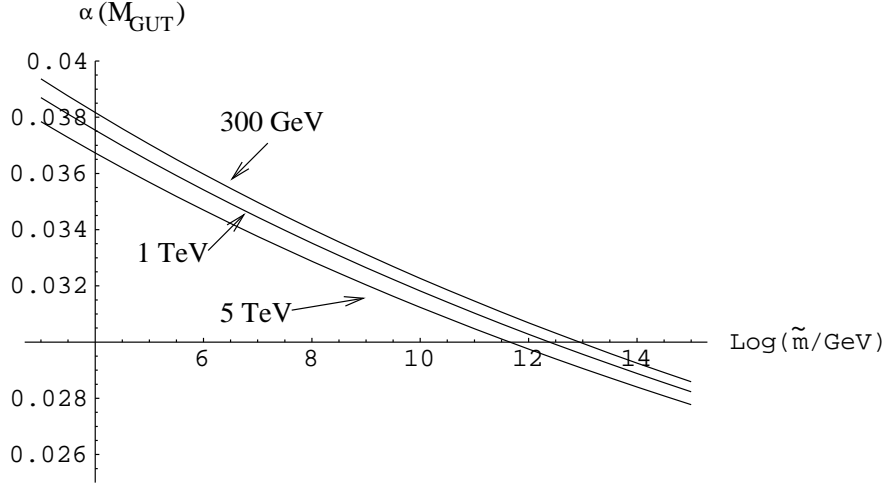}
\caption{In the case of Gaugino mass unification at scale
$\tilde{m}$, we plot the prediction for $\alpha_s(M_{GUT})$. The
results for $M_2=300$ GeV, 1 TeV and 5 TeV are shown.}
\label{alphaGut_Unification}
\end{center}
\end{figure}

\begin{figure}[!h]
\begin{center}
\includegraphics[scale=0.75]{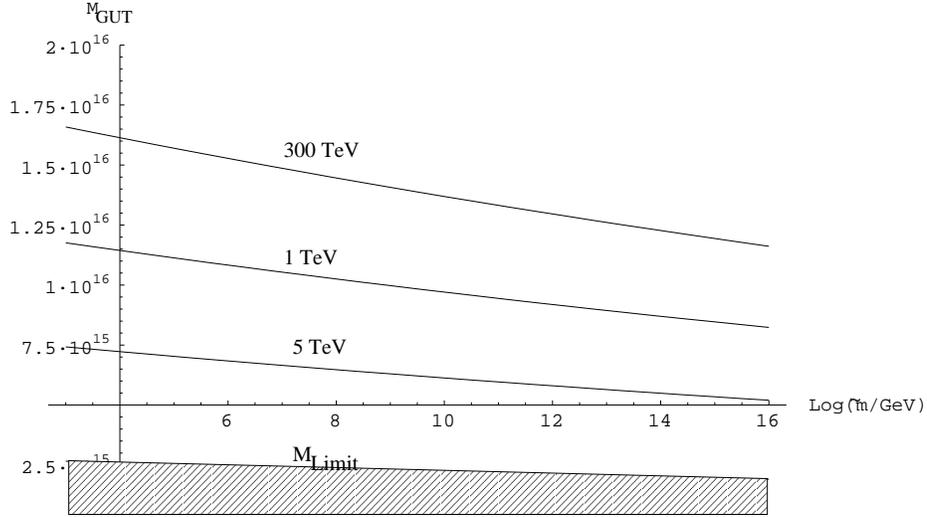}
\caption{In the case of Gaugino mass unification at scale
$\tilde{m}$, we plot the prediction for $M_{GUT}$. The results for
$M_2=300$ GeV, 1 TeV and 5 TeV are shown, together with the lower
bound on $M_{GUT}$ from Proton decay.} \label{Mgut_Unification}
\end{center}
\end{figure}

As we will see later, a particular interesting case for the LSP in
the hidden sector is given by a Bino NLSP. For this case, we need
to do a more accurate computation, splitting the mass of the
Gauginos, from that of the
Higgsinos, and taking the Wino mass roughly two times larger than
the Bino mass, as inferred from \cite{Giudice:2004tc} 
for gaugino mass unification initial conditions. In
fig.\ref{Gut_Unification_Higgsino_fig}, we show what is the
allowed region for the mass of the Bino and the ratio of the
Hissino mass and Bino mass, such that gauge coupling unification
is attained with a mass for the scalars, $\tilde{m}$, in the range
$10^5$ GeV-$10^{18}$ GeV. Raising the Higgsino mass with respect
to the Bino mass has the effect of lowering the maximum mass for
the fermionic superpartners. This is due to the fact that, raising
the Higgsino mass, the unification value for the U(1) and SU(2)
couplings is reduced, so that the prediction for $\alpha_s(M_Z)$
is lowered.

\begin{figure}[!h]
\begin{center}
\includegraphics{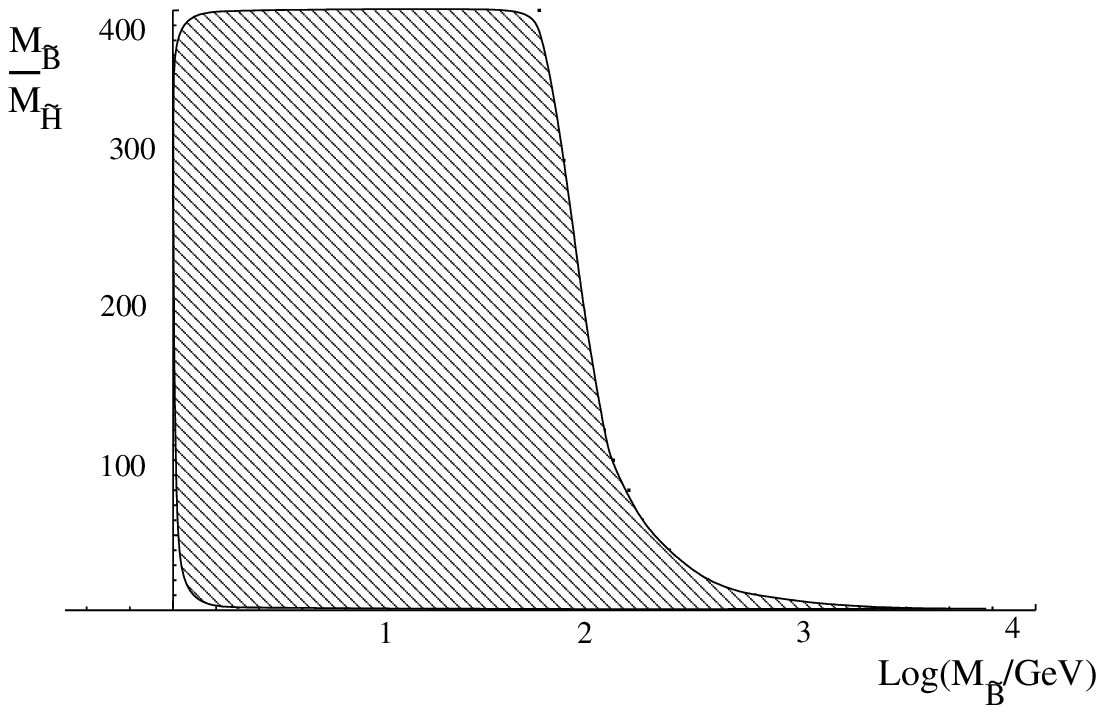}
\caption{Shaded is the allowed region for the Bino mass and the
ratio of the Higgsino mass and the Bino mass, in order to obtain
Gauge Coupling Unification with a value of the scalar mass
$\tilde{m}$ in the range $10^5$ GeV-$10^{18}$ GeV. We take
$M_2\simeq 2 M_1$ as inferred from gaugino mass unification at the GUT scale
\cite{Giudice:2004tc}}
\label{Gut_Unification_Higgsino_fig}
\end{center}
\end{figure}

\subsection{Gaugino Mass Condition from Anomaly Mediation}

Of the possible initial conditions for the Gaugino mass which can
have some influence on the upper bound on fermions mass, there is
one which is particularly natural, and which is coming from
Anomaly Mediated Susy breaking, and according to which the initial
conditions for the gaugino masses are:
\begin{equation}
M_i=\frac{\beta_{g_i}}{g_i}m_{3/2}\sim \frac{c_i g^2_i}{16\pi^2}
m_{3/2}
\end{equation}
where $\beta_i$ is the beta-function for the gauge coupling, and
$c_i$ is an order one number. These initial conditions are not
relevant for the Gravitino LSP, as in this case the Neutralinos
are lighter than the Gravitinos; but they can be relevant in the
case the LSP is given by a fermion in the hidden sector, as we
will study later. Further, the study of this case is interesting
on its own, as it gives an upper
bound on the fermionic superpartners which is higher with respect to
the one coming from gaugino mass unification initial conditions.\\

The study parallels very much what done in the former section, with
the only difference being the fact that in this case, as computed
in \cite{Giudice:2004tc}, the mass hierarchy between the Gluinos
and the Gauginos is higher ( a factor $\sim 10-20$ instead of
$\sim 3-8$). This has the effect of raising the allowed mass for
the fermions. We do the same amount of translation as before for
the predicted $\alpha_s(M_Z)$. The result is shown in
fig.\ref{Gut_Anomaly}, and gives, as upper limit, $M_2=18$
TeV.

\begin{figure}[!h]
\begin{center}
\includegraphics[scale=0.85]{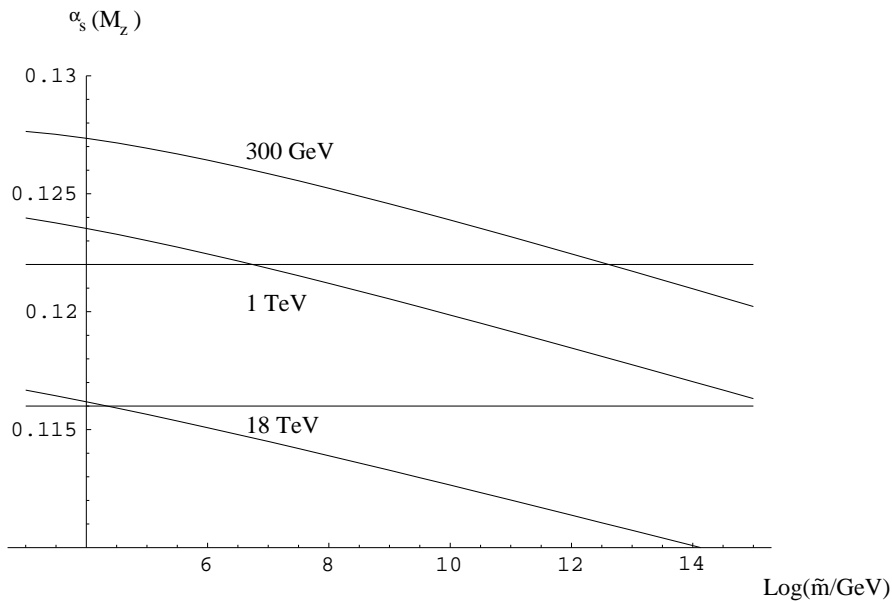}
\caption{In the case of Gaugino mass condition from anomaly
mediation at scale $\tilde{m}$, we plot the unification prediction
for $\alpha_s(M_Z)$. The results for $M_2=300$ GeV, 1 TeV and 18
TeV are shown. The horizontal lines represent the experimental
bounds} \label{Gut_Anomaly}
\end{center}
\end{figure}

In fig.\ref{Mgut_Anomaly}, and fig.\ref{alphaGut_Anomaly}, we plot
the predictions for $\alpha_{GUT}$ and $M_{GUT}$ for the same
range of masses, and we see that unification is reached in the
perturbative regime, and that the unification scale is large enough to
avoid proton decay limits, but it is getting very close to the
experimental
bound for large values of the mass $\tilde{m}$.\\

\begin{figure}[!h]
\begin{center}
\includegraphics{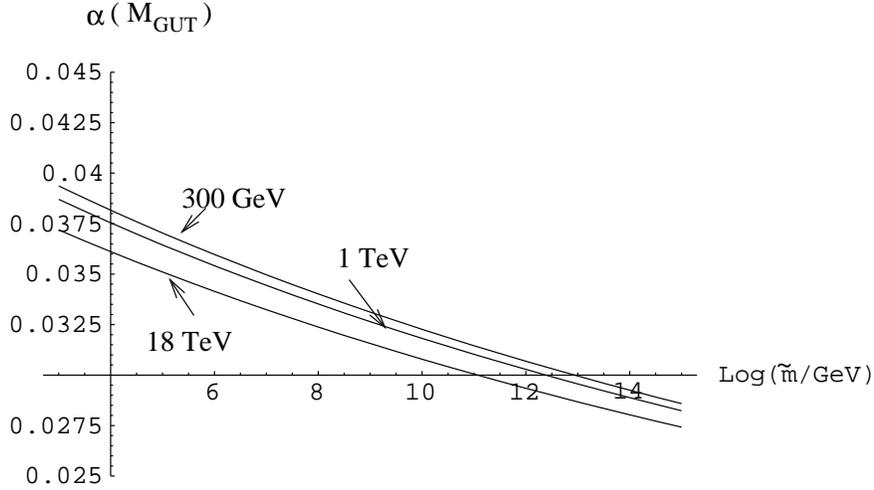}
\caption{In the case of Gaugino mass from Anomaly Mediation,
we plot the prediction for $\alpha_{GUT}$. The
results for $M_2=300$ GeV, 1 TeV and 18 TeV are shown.}
\label{alphaGut_Anomaly}
\end{center}
\end{figure}

\begin{figure}[!h]
\begin{center}
\includegraphics[scale=0.75]{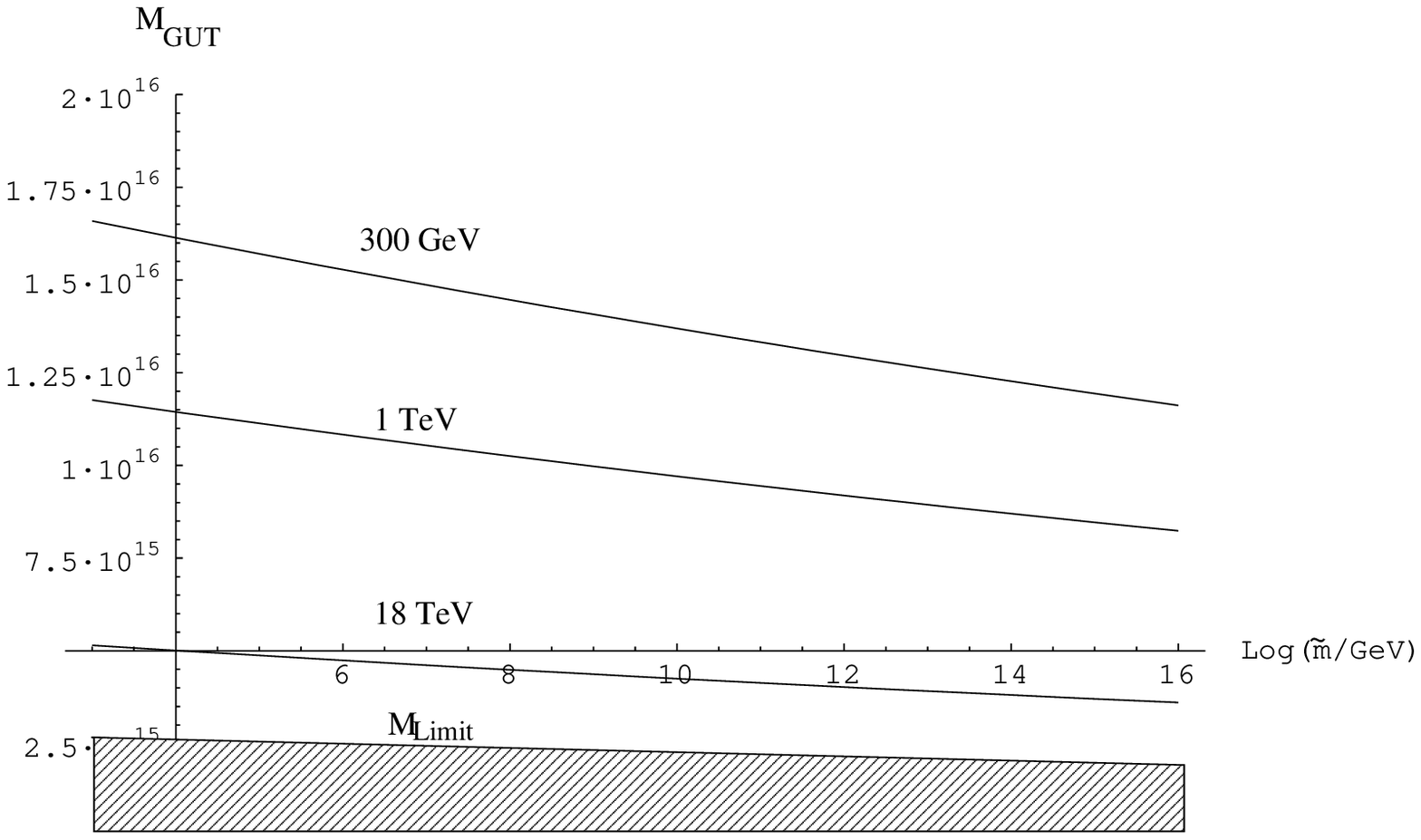}
\caption{In the case of Gaugino mass from Anomaly Mediation,
we plot the prediction for $M_{GUT}$. The results for
$M_2=300$ GeV, 1 TeV and 18 TeV are show, together with the lower
bound on $M_{GUT}$ from Proton decay. } \label{Mgut_Anomaly}
\end{center}
\end{figure}

As we can see, in the case of Gaugino Mass from Anomaly Mediation,
the upper limit on fermion mass is raised to 18 TeV.
This last one can be interpreted as a sort of maximum allowed mass
for fermionic superpartners.\\

It is important to note that, as pointed out in
\cite{Giudice:2004tc}, in this case the Bino can not be the NLSP.

\section{Hidden sector LSP and Dark Matter Abundance}

An hidden sector LSP which is very weakly interacting can well be
the DM from the astrophysical and cosmological point of view. Its
present abundance can be given by two different sources: it can be
a thermal relic, if in the past the temperature was so high that
hidden sector particles were in equilibrium with the thermal bath,
or it can be present in the universe just as the
result of the decay of the other supersymmetric particles.\\

We concentrate in the case in which the thermal relic abundance is
negligible, which is generically the case for not too large reheating
temperatures, and the abundance is given by the decaying of the
other supersymmetric particles into the LSP. A discussion on the
consequences of a thermal relic abundance of Gravitinos is
discussed
in \cite{Arkani-Hamed:2004yi}.\\

In our case, the relic abundance of the heavier particles is what
determines the final abundance of the LSP, and so it is the
fundamental quantity to analyze. In the very early universe, the
typical time scale of the cosmic evolution $H^{-1}$ is much larger
than the time scale of interaction of a weakly interacting
particle, and so a weakly interacting particle is in thermal
equilibrium. Therefore, its abundance is given by the one of a
thermal distribution. As the temperature of the universe drops
down, the interaction rate is not able anymore to keep the
particle in thermal equilibrium, and so the particle decouples
from the thermal bath, and its density begins to dilute, ignoring
the rest of the thermal bath. We say in this case that the
particle species
freezes out.\\

In the case of weakly interacting particles around the TeV scale,
the freeze out temperature is around decades of GeV, and so they
are non relativistic at the moment of freezing out. In this case,
the relic abundance of these particles is given by the following
formula \cite{Bernstein:1985th,Scherrer:1985zt,Kolb}:
\begin{equation}
\Omega_{NLSP}h^2\approx0.1 \left(\frac{10^{-9} {\rm
GeV}^{-2}}{<\sigma
v>}\right)\left(\frac{15}{\sqrt{g_*}}\right)\left(\frac{10^{19}
{\rm GeV}
}{M_{pl}}\right)\left(\frac{x_f}{30}\right)\left(\frac{h^2}{0.5}\right)
\label{Omega_NLSP}
\end{equation}
where $<\sigma v>$ is the thermally averaged cross section at the
time of freeze out, $x_f=\frac{m_{NLSP}}{Tf}$ where $T_f$ is the
freeze out temperature, $g_*$ is the effective number of degrees
of freedom at freeze out, and $h$ is the
Hubble constant measured in units of $100 {\rm Km}/({\rm sec}\ {\rm Mpc})$.\\
It is immediate to see that, for weakly interacting particle at 1
TeV, the resulting $\Omega$ is of order unity, and this has led to
the claim that the Dark Matter bounds some supersymmetric partners
to be at TeV scale. In this paper, we shall check this claim
for an LSP in the hidden sector.\\

Once the weakly interacting particles are freezed out, they will
rapidly decay to the NLSP, which, being lighter, will be in
general still in thermal equilibrium. So, it will be the NLSP the
only one to have a relevant relic abundance, determined by the
freeze out mechanism, and so it will be the NLSP that, through
its decay, will generate the present abundance of the LSP.\\

In Split Susy, the NLSP can either be the lightest Neutralino, or
the lightest Chargino. The Neutralino is a mixed state of the
interaction eigenstates Bino, Wino, and neutral Higgsinos, and is
the lightest eigenstate of the following
matrix\cite{Pierce:2004mk}:
\begin{equation}
\left(\begin{array}{cccc}M_1&0&-\frac{k'_2v}{\sqrt{8}}&\frac{k'_1v}{\sqrt{8}}\\
0&M_2&\frac{k_2v}{\sqrt{8}}&-\frac{k_1v}{\sqrt{8}}\\
-\frac{k'_2v}{\sqrt{8}}&\frac{k_2v}{\sqrt{8}}&0&-\mu\\
\frac{k'_1v}{\sqrt{8}}&-\frac{k_1v}{\sqrt{8}}&\mu&0\end{array}\right)\
\label{neutralino_matrix}
\end{equation}
which differs from the usual Neutralino matrix in low energy Susy
for the Yukawa coupling, which have their Susy value at the Susy
breaking scale $\tilde{m}$, but then run differently from that scale to the
weak scale.\\

The Chargino is a mixed eigenstate of charged Higgsino and charged
Wino, and is the lightest eigenstate of the following matrix:
\begin{equation}
\left(\begin{array}{cc}M_2&\frac{k_1v}{2}\\
\frac{k_2v}{2}&\mu\end{array}\right)\ \label{chargino_matrix}
\end{equation}

The actual and precise computation of the thermally averaged cross
section of the NLSP at freeze out, which determines the
$\Omega_{NLSP}$, is very complicated, because there are many
channels to take care of, which depend on the abundance of the
particles involved, and on the mixing of states, creating a very
complicated system of differential equations. A software called
DarkSusy has been developed to reliably compute the relic
abundance \cite{Gondolo:2002tz}, and, in a couple of recent papers
\cite{Giudice:2004tc,Pierce:2004mk}, it has been used to compute
the relic abundance of the Neutralino NLSP in Split Susy. In this
kind of theories, in particular, this computation is a bit
simplified, since the absence of the scalar superpartners makes
many channels inefficient. Nevertheless, in most cases, the
computation is still too
complicated to be done analytically.\\

In this paper, we consider both the possibility that the NLSP is a
Neutralino and a Chargino.\\
In the case of Neutralino NLSP, we modify the Dark Susy code  
\cite{Gondolo:2002tz} and adapt it to the case of Split Susy. 
We consider the cases of pure Bino, pure Wino,
pure Higgsino, and Photino NLSP. In the case of
Chargino NLSP, we consider the case of charged Higgsino and
Charged Wino as NLSP, and we estimate their abundance with the
most important diagram. We will see, in fact, that in this case a
more precise determination of the relic abundance is
not necessary.\\

Once the NLSP has freezed out, it will dilute for a long time,
until, at the typical time scale of 1 sec, it will decay
gravitationally to the LSP, which will be stable, and will
constitute today's Dark Matter. It's present abundance is
connected to the NLSP "would be" present abundance by the simple
relation:

\begin{equation}
\Omega_{LSP}=\frac{m_{LSP}}{m_{NLSP}}\Omega_{NLSP}\label{Omega_LSP}
\end{equation}

Already from this formula, we may get some important information
on the masses of the particles, just comparing with the case in
which the Neutralino or the Chargino is the LSP. In fact, since
$\frac{m_{LSP}}{m_{NLSP}}<1$, $\Omega_{NLSP}$ has to be greater
than what it would have to have if the NLSP was the LSP, in order for the
LSP to provide all the DM. The
abundance of the NLSP is inversely proportional to $<\sigma v>$,
and this means that we need to have a typical cross section
smaller than the one we would obtain in the case of a weakly
interacting particle at TeV scale. This result can be achieved in
two ways: either raising the mass of the particles, since
$\sigma\sim\frac{1}{m^2}$, or by choosing some particle which for
some reason is very low interacting.\\

The direction in which the particle become very massive is not
very attractive from the LHC detection point of view, but still,
in Split Susy, is in principle an acceptable scenario.\\

The other direction instead immediately lets a new possible candidate to
emerge, which could be very attractive from the LHC detection point of
view. In fact, in Split Susy, a pure Bino NLSP is almost not
interacting, the only annihilation channel being the one into
Higgs bosons in which an Higgsino is exchanged. In this case the
relic abundance has $\Omega_{NLSP}\gg1$, and this was the reason why a
pure Bino could not be the DM in Split Susy
\cite{Giudice:2004tc,Pierce:2004mk}. In the case of a
gravitationally interacting LSP, as we are considering, this over
abundance would go exactly into the right direction, and 
it could open a quite interesting region for detection at LHC.\\

\section{Cosmological Constraints}

Since we wish the LSP to be the Dark Matter of the universe, so,
we impose its abundance to cope with WMAP data
\cite{Bennett:2003bz,Spergel:2003cb}.\\

In general, for low reheating temperature, only the weakly
interacting particles are thermally produced, and only the NLSP
will remain as a thermal relic in a relevant amount. Later on, it will decay
to the LSP. This decay will give the strongest cosmological
constraints.\\

In fact, concentrating on the Gravitino, which interacts only
gravitationally, we can naively estimate its lifetime as:
\begin{equation}
\Gamma\sim\frac{m^3_{NLSP}}{M^2_{pl}}
\end{equation}
where the $m_{NLSP}$ term comes from dimensional analysis. In
reality, we can easily do better. In fact, as we have Goldstone
bosons associated to spontaneous symmetry breaking, the breaking
of supersymmetry leads to the presence of a Goldstino, a massless
spinor. Then, as usually occurs in gauge theories, the Goldstino
is eaten by the massless Gravitino, which becomes a massive
Gravitino with the right number of polarization. Therefore, the
coupling of the longitudinal components of the Gravitino to the
LSP will be determined by the usual pattern of spontaneous
symmetry braking, and in particular will be controlled by the
scale of symmetry
breaking. This means that the coupling constant may be amplified.\\
In fact, if we concentrate on the Gauginos for simplicity, we can
reconstruct their coupling to the Goldstino simply by looking at
the symmetry braking term in the lagrangian in unitary gauge, then
reintroducing the Goldstino performing a Susy transformation, and
promoting the transformation parameter to a new field, the
Goldstino. The actual coupling is then obtained after canonical
normalization of the Goldstino kinetic term, which is obtained
performing a Susy transformation of the mass term of the
Gravitino, and remembering that in the case of Sugra, the Susy
transformation of the Gravitino contains a piece proportional to
the vacuum energy. In formulas, the Gaugino Susy transformation is
given by:
\begin{equation}
\delta \lambda=\sigma^{\mu\nu} F_{\mu\nu}\xi
\end{equation}
where $\xi$ is the Goldstino. This implies that the mass term of
the Gaugino sources the following coupling between the Gaugino and
the Goldstino:
\begin{equation}
\delta (m\lambda\lambda)\supset m_\lambda \lambda
\sigma^{\mu\nu}F_{\mu\nu}\xi
\end{equation}
The Goldstino kinetic term cames from the Gravitino tranformation,
which is:
\begin{equation}
\delta \psi_\mu=m_{Pl}\partial_\mu\xi+i f \sigma_\mu \bar\xi
\end{equation}
so, the Gravitino mass term produces the Goldstino kinetic
term:
\begin{equation}
\delta (m_{gr}\psi_\mu \sigma^{\mu\nu}\psi_\nu)\supset m_{gr}f
m_{pl}\bar{\xi}\sigma^\mu\partial_{\mu}\xi=f^2
\bar{\xi}\sigma^\mu\partial_{\mu}\xi
\end{equation}
where in the last expression we used that $m_{gr}=f/m_{Pl}$. So,
after canonical normalization, we get the following interaction
term:
\begin{equation}
L_I=\frac{m_\lambda}{f^2}\lambda \sigma^{\mu\nu}F_{\mu\nu}\xi_c
\end{equation}
where $\xi_c=\xi f$ is the canonically normalized Goldstino. After
all this, we get an enhanced decay width like this:
\begin{equation}
\Gamma\sim \frac{1}{M^2_{pl}}\left(\frac{m_{NLSP}}{m_{gr}}\right)^2m^3_{NLSP}
\label{estimate}
\end{equation}
Note that this is independent on the particle species, as it must
be by the equivalence principle \cite{Cyburt:2002uv}.\\

Plugging in some number, we
immediately see that, for particles around the TeV scale, without
introducing a big hierarchy with the Gravitino, the time
of decay is approximately $\sim 1$
sec, and this is right the time of Big Bang Nucleosynthesis (BBN).\\
This is the origin of the main cosmological bound. In fact, the
typical decay of the LSP will be into the Gravitino and into its
SM partner. The SM particle will be very energetic, especially
with respect to a thermal bath which is of the order of 1 MeV, and
so it will create showers of particles, which will destroy some of the
existent nuclei, and enhance the formation of others, with the
final result of alterating the final abundance of the light
elements \cite{Cyburt:2002uv}.\\

There is also another quantity which comes into play, and it is
what we can call the "destructive power". In fact, the alteration
of the light nuclei abundance will be proportional to the product
of the abundance of the decaying particle and to the energy
release per decay. This information is synthesized 
in an upper limit on the variable
$\xi$ defined as:
\begin{equation}
\xi=B \epsilon Y
\end{equation}
where $B$ is the branching ratio for hadronic or electromagnetic
decays (it turns out that hadronic decays impose constrains a
couple of order of magnitude more stringent that electromagnetic
decays), $\epsilon$ is the energy release per decay, and finally
$Y=\frac{n_{\chi}}{n_s}$, where $n_s$ is the number of photons per
comoving volume, and $n_\chi$ the number of decaying particles per
comoving volume. Again, it is easy to see what will be the lower limit
on the upper limit on $\xi$. For the moment, we will neglect the dependence
on the branching ratio $B$, because, clearly, one of the two branching
ratios must be of order one. Then, we understand that the most dangerous
particles for BBN will be those particles that decay
when BBN has already produced most of the nulcei we have to see today.
A particle which decays earlier than this time, will in general
have its decay products diluted and thermalized with an efficiency
that depends on the kind of decay product of the particle: either
baryonic or electromagnetic, and it turns out that the dilution
for electromagnetic decays is much more efficient. So, it is clear that the
upper limit on the ``desctructive power'' $\xi$ will be lower for particles 
which decay after BBN. For these late decaying particles, we can estimate
what the upper limit on $\xi$ should be with the following argument.
$\xi$ will become dangerous if the
energy release is bigger than 1 MeV, in order for the decay product 
to be able to
destroy nuclei, and also if $Y$ is greater than $\frac{n_B}{n_s}$
which represent the number of baryons per comoving volume
opportunely normalized. Plugging in the numbers, with, again, $B\sim 1$, we get
\begin{equation}
\xi_{dangerous}\gtrsim 10^{-14} {\rm GeV} \label{xi_dangerous}
\end{equation} 
This values of $\xi_{dangerous}$  is more or less where the limit
seems to apply in numerical simulations for late decaying particles, 
and it is in fact independent 
on the particular kind of decay, as in this case there are not dilution
issues \cite{Kawasaki:2004yh,Kawasaki:2004qu}. For
early decaying particles the limits do depend on the kind of decay, and they 
get more and more relaxed as the decay time becomes shorter and shorter, 
until there is practically no 
limit on particles which decay earlier than $\sim 10^{-2}$ sec.
Notice however that, from the estimates above on the decay time, the particles
which we will be interested in will tend to decay right in the region where 
these limits apply.\\
The limit in eq.(\ref{xi_dangerous}) translates into
another useful parameter:
\begin{equation}
\Omega_{dangerous}\gtrsim 10^{-7}
\end{equation}
for the contribution of the NLSP around the time of
nucleosynthesis. An easy computation shows that, imposing
$\Omega_{DM}\sim 1$ today, we get that the contribution of NLSP
goes as:
\begin{equation}
\Omega_{NLSP}\sim 10^{-7} \frac{m_{NLSP}}{m_{gr}} \left(\frac{{\rm
MeV}}{T}\right)
\end{equation}
This estimates are obviously very rough, but they are useful to
give an idea of the physics which is going by, and they are, at
the end of the day, quite accurate. They nevertheless tell us that
we are really in the region in which these limits are effective,
with two possible consequences: on one hand, a big part of the
parameter region might be excluded, but also, on the other hand, this
tells us that a possible indirect detection through
deviations from the standard picture
nucleosynthesis might reveal new physics.\\

In two recent papers \cite{Kawasaki:2004yh,Kawasaki:2004qu},
numerical simulation were implemented to determine the constraints
on $\xi$, both for the hadronic and the electromagnetic decays,
and we shall use
their data. (See also  
 \cite{Jedamzik:2004ip,Jedamzik:2004er}
where a similar discussion is developed.)\\

Cosmological constraints come also from another observable. A late
decaying particle can in fact alter the thermal distribution of
the photons which then will form the CMB, introducing a chemical
potential in the CMB thermal distribution bigger than the one
which is usually expected due to the usual cosmic evolution, or
even bigger than the current experimental upper
bound \cite{Fixsen:1996nj,Hagiwara:2002fs}.\\
Analytical formulas for the produced effect are given in
\cite{Hu:1993gc,Hu:1992dc}. Nevertheless, it is useful to notice
that the induced chemical potential $\mu$ is mostly and hugely
dependent on the time of decay of the particles. In particular, we
see that:
\begin{equation}
\mu\sim e^{-\frac{\tau_{dc}}{\tau_{NLSP}}}
\end{equation}
where $\tau_{NLSP}$ is the lifetime of the NLSP, and
$\tau_{dc}\sim 10^6$s is the time at which the double Compton
scattering of the photons is no more efficient. The
$\xi_{dangerous}$ for this quantity is $\xi_{dangerous}\sim
10^{-9}$ GeV. So, we conclude that basically, for
$\tau_{NLSP}<\tau_{dc}$ , there are no limits, while for
$\tau_{NLSP}>\tau_{dc}$, the limit from nucleosynthesis is
stronger. We easily see that this constraint never comes into
play in our work.\\

From formula (\ref{estimate}), we can already extract an idea of
what will be the final result of the analysis. In fact, we can
avoid the limits from nucleosynthesis by decaying early. This
means that, according to (\ref{estimate}), we need to let the
ratio $\frac{m_{NLSP}}{m_{LSP}}$ to grow, and consequently
$\Omega_{NLSP}$ has to grow as well. This leads to two directions:
either a very massive LSP or a a very weakly interacting LSP. The
first direction goes in agreement with one of the directions we
had found in order to match the constraint from $\Omega_{DM}$, and
tells us that, in general, a massive NLSP will be acceptable from
the cosmological point of view. However, it will have chanches to
encounter the constraints coming from gauge couplings unification.
The other direction is to have an NLSP whose main annihilation
channel is controlled by another particle, which can be made
heavy. As an example, this is the case for the Bino, whose channel
is controlled by the Higgsino: so, we might have a light Bino, if
the Higgsino will be heavier.\\

\section{Gravitino LSP}

In this section, we concentrate in detail on the possibility that
the LSP in Split Susy is the Gravitino, and that it constitute the
Dark Matter of the universe. We shall consider the mass of the
Gravitino as a free parameter, and we shall try to extract
information on the mass and the nature of the NLSP. However, an
actual lower limit on the Gravitino mass can be expected in the
case Susy is broken directly, as in that case the mass of the
Gravitino should be: $m_{gr}\sim\frac{\tilde{m}^2}{M_{pl}}$, where
$\tilde{m}$ is the Susy breaking scale. Since, roughly, in Split
Susy $\tilde{m}$ is as light as $\sim 100$ TeV, we get the lower
limit $m_{gr}\gtrsim 10^{-8}$ GeV, which, as we will see, is lower
than the region we will concentrate on.\\

As we learnt in the former two sections, there are two fundamental
quantities to be computed: the lifetime of the NLSP, and
$\Omega_{NLSP}$.\\

 As we said before, we shall consider both the
Neutralino and the Chargino as LSP. The decaying amplitude of a
Neutralino into Gravitino plus a Standard Model particle was
computed in \cite{Feng:2004mt,Feng:2003xh,Feng:2003uy}.\\

For decay into Photons:

 \begin{equation}
\Gamma(\chi\rightarrow\gamma,gr)=\frac{|N_{11}\cos(\theta_w)+N_{12}\sin(\theta_w)|^2}{48\pi
M^2_{pl}}
\frac{m^5_\chi}{m^2_{gr}}\left(1-\frac{m^2_{gr}}{m^2_\chi}\right)^3
\left(1+3\frac{m^2_{gr}}{m^2_\chi}\right)
\label{photon_decay}
 \end{equation}

where $\chi=N_{11}(-i\tilde
B)+N_{12}(-i\tilde{W})+N_{13}\tilde{H_d}+N_{14}\tilde{H_u}$ is the
NLSP. As we see, eq.(\ref{estimate}) reproduces the right behavior
in the limit $m_{NLSP}/m_{LSP}\gg1$ . This decay will contribute
only to ElectroMagnetic (EM)
energy.\\

The leading contribution from hadronic decays comes from the decay
into $Z,gr$ and $h,gr$. These decays will contribute to EM or
Hadronic energy according to the branching ratios of the SM
particles.\\
The decay width to Z boson is given by:
 \begin{eqnarray}
\Gamma(\chi\rightarrow
Z,gr)=&&\frac{|-N_{11}\sin(\theta_w)+N_{12}\cos(\theta_w)|^2}{48\pi
M^2_{pl}} \frac{m^5_\chi}{m^2_{gr}}F(m_\chi,m_{gr},m_z)\\
\nonumber
&&\left(\left(1-\frac{m^2_{gr}}{m^2_\chi}\right)^2\left(1+3\frac{m^2_{gr}}
{m^2_{\chi}}\right)
-\frac{m^2_Z}{m^2_{gr}}G(m_\chi,m_{gr},m_Z)\right)
 \label{Z_decay}
 \end{eqnarray}
where
\begin{equation}
F(m_\chi,m_{gr},m_z)=\left(\left(1-\left(\frac{m_{gr}+m_Z}{m_\chi}\right)^2
\right)\left(1-\left(\frac{m_{gr}-m_Z}{m_\chi}\right)^2\right)\right)^{1/2}
\end{equation}
\begin{equation}
G(m_\chi,m_{gr},m_z)=3+\frac{m^3_{gr}}{m^3_\chi}\left(-12
+\frac{m_{gr}}{m_\chi}+\frac{m^4_Z}{m^4_\chi}-\frac{m^2_Z}{m^2_\chi}
\left(3-\frac{m^2_{gr}}{m^2_\chi}\right)\right)
\end{equation}
The decay width to $h$ boson is given by:
 \begin{eqnarray}
\Gamma(\chi\rightarrow
h,gr)=&&\frac{|-N_{13}\sin(\beta)+N_{14}\cos(\beta)|^2}{48\pi
M^2_{pl}} \frac{m^5_\chi}{m^2_{gr}}F(m_\chi,m_{gr},m_z)\\
\nonumber
&&\left(\left(1-\frac{m^2_{gr}}{m^2_\chi}\right)^2\left(
1+\frac{m^2_{gr}}{m^2_{\chi}}\right)^4
-\frac{m^2_h}{m^2_{gr}}H(m_\chi,m_{gr},m_h)\right)
 \label{h_decay}
 \end{eqnarray}
where $h=-H^0_d sin(\beta)+H^0_uCos(\beta)$ is the fine tuned
light Higgs, and
\begin{equation}
H(m_\chi,m_{gr},m_h)=3+4\frac{m_{gr}}{m_\chi}+2\frac{m^2_{gr}}{m^2_\chi}+4\frac{m^3_{gr}}{m^3_{\chi}}+3\frac{m^4_{gr}}{m^4_\chi}
+\frac{m^4_h}{m^4_\chi}-\frac{m^2_h}{m^2_\chi}
\left(3+2\frac{m_{gr}}{m_\chi}+3\frac{m^2_{gr}}{m^2_\chi}\right)
\end{equation}

We further use the following branching ratios and energy release parameters:
\begin{equation}
B^\chi_{EM}\sim 1
\end{equation}
\begin{equation}
\epsilon^\chi_{EM}=\frac{m^2_\chi-m^2_{gr}}{2m_\chi}
\end{equation}
\begin{equation}
B^\chi_{Had}\sim\frac{\Gamma(\chi\rightarrow Z,gr)
B^Z_{had}+\Gamma(\chi\rightarrow h,gr)
B^h_{had}+\Gamma(\chi\rightarrow
q,\bar{q},gr)}{\Gamma(\chi\rightarrow
\gamma,gr)+\Gamma(\chi\rightarrow h,gr)+\Gamma(\chi\rightarrow
Z,gr)}
\end{equation}
\begin{equation}
\epsilon^\chi_{had}=\frac{m^2_\chi-m^2_{gr}-m^2_{Z,h}}{2m_\chi}
\end{equation}
where $\epsilon^\chi_i$ is the energy release per decay in the EM
and in the Hadronic channel, and $B^\chi_i$ is the branching ratio
in the EM and Hadronic channel. We use $B^h_{had}\sim
0.9,B^Z_{had}\sim 0.7$. Since it will not play an important role,
we just estimate the channel $\Gamma(\chi\rightarrow
q,\bar{q},gr)$, and do not perform a complete computation. This
channel provides the hadronic decays when $m_{NLSP}-m_{gr}$ is
less than the $m_Z$ or $m_h$. The leading diagram in this case is
given by the tree level diagram in which there is a virtual $Z$
boson or a virtual Higgs that decays into quarks.

\subsection{Neutral Higgsino, Neutral Wino, and Chargino NLSP}

An Higgsino NLSP will be naturally much interacting in Split Susy, 
quite independently of the
other partners mass. In fact, there are gauge interaction and
Yukawa coupling to the other particles. While the coupling to the
Z vanishes for $\mu\gg m_{Z}$, in that case neutral Higgsino and
Charged Higgsino are almost degenerate, and so the interaction
with the Higgs become relevant. This means that the annihilation
rate will never be very weak, and so $\Omega_{NLSP}$ will be large
only for large $\mu$. An analytical computation is too complicated
for our necessities, even with the simplifications of Split Susy,
so, we modify the DarkSusy code \cite{Gondolo:2002tz} 
to adapt it to the Split Susy case,
and we obtain the following relic abundance:
\begin{equation}
\Omega_{\tilde{H^0}}h^2=0.09 \left(\frac{\mu}{{\rm TeV}}\right)^2
\end{equation}
In order to avoid nucleosynthesis constraints, we need to decay
early. This can be achieved either raising the hierarchy between
Higgsino and Gravitino,
or raising the mass of the Higgsino. Since
$\Omega_{LSP}=\frac{m_{LSP}}{m_{NLSP}}\Omega_{NLSP}$, we can
not grow too much with the hierarchy, and so we are forced to
raise the mass of the Higgsino.\\

This is exactly one of the two directions to go in the
parameter space we had outlined in the first sections, and it is the one
which is less favourable for detection at LHC.\\

The results of an actual computation are shown in
fig.\ref{Higgsino}, where  we plot the allowed
region for the Higgsino NLSP, in the plane $m_{gr},\delta
m=m_{NLSP}-m_{gr}$. The quantity $\delta m$ well represents the
available energy for decay, and, obviously, can not be negative. \
The Hadronic and the Electromagnetic constraints
we use come from the numerical simulations done in 
\cite{Kawasaki:2004yh,Kawasaki:2004qu}. There, constraints are given as upper
limit on the quantity  $\xi=B_{(EM,Had)} \epsilon_{(EM,Had)} Y$ as a function 
of the time of decay. We then apply this limit to our NLSPs computing both
the time of decay and the quantity $\xi$ with the formulas given in the 
former section. \\

\begin{figure}[!h]
\begin{center}
\includegraphics[scale=1.1]{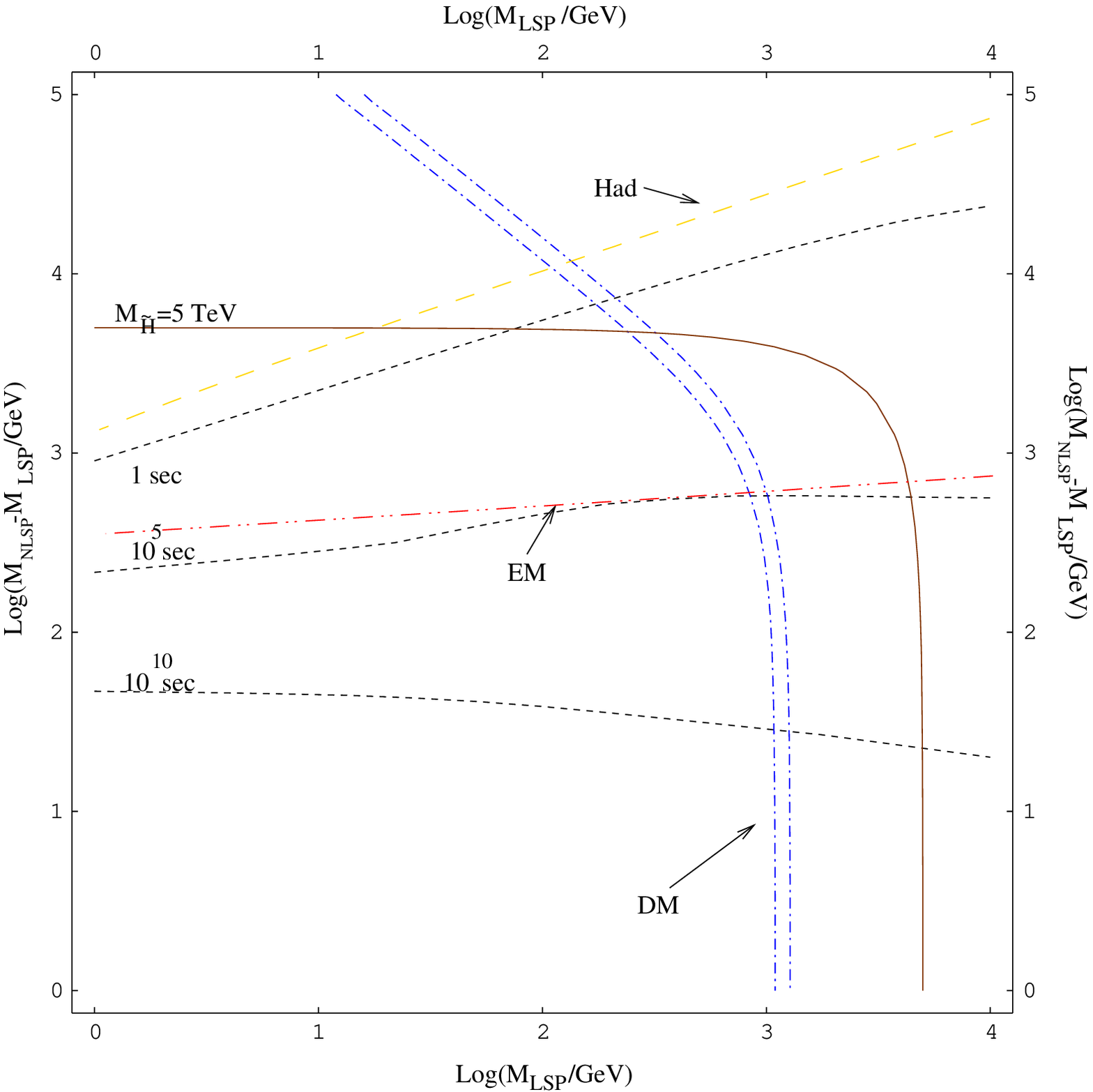}
\caption{Constraints for the Higgsino NLSP, Gravitino LSP. There
is no allowed region. The long dashed contour delimitats from above the
excluded region by the hadronic constraints from BBN,
the dash-dot-dot contour represents the same for EM constraints from
BBN, the dash-dot lines represent the region within which
$\Omega_{DM}$ is within the experimental limits; finally, we show
the short dashed countors where the Higgsino decays to Gravitino at 1
sec, $10^5$ sec, and $10^{10}$ sec, and the solid line where
the Higgsino is 5 TeV heavy, which represents the upper bound for
Gauge Coupling Unification. For Neutral Wino and Chargino NLSP,
the result is very similar. }\label{Higgsino}
\end{center}
\end{figure}

As we had anticipated, the cosmologically allowed region is given by:
\begin{equation}
m_{gr}\leq 4\times 10^2{\rm GeV}
\end{equation}
\begin{equation}
m_{\tilde{H}}\geq 20 {\rm TeV}
\end{equation}
This mass range is not allowed by gauge coupling unification, as,
for Gravitino LSP, we have to use the upper bound on NLSP coming
from gaugino mass unification initial conditions. So we conclude
that the Higgsino NLSP is excluded. This is a very nice example of
how much we can constraint physics combining
particle physics data, and cosmological observations.\\

Finally, note how the hadronic constraints raise the limit on the
Higgisino mass
of approximately one order of magnitude.\\

In the case of the Neutral Wino NLSP, and Chargino NLSP, there are
basically no relevant differences with respect to the case of the
Neutral Higgsino, the main reason being the fact that the many
annihilations channels lead to an high mass for the NLSP, exactly
in parallel to the case of the Higgsino. We avoid showing
explicitly the results, and simply say that, for all of them, the
cosmologically
allowed parameter space is very similar in shape and values to the
one for Higgsino, with just this slight correction on the
numerical values:
\begin{equation}
m_{gr}\leq 5\ 10^2{\rm GeV}
\end{equation}
\begin{equation}
m_{\tilde{W}^0}\geq 30 {\rm TeV}
\end{equation}
for the Wino case. Notice that the mass is a little higher than in
the Higgsino case, as the Wino is naturally more interacting. In
fact, its relic abundance is given by (again, using a on porpuse modified
version of Dark Susy \cite{Gondolo:2002tz} ):
\begin{equation}
\Omega_{\tilde{W^0}}h^2=0.02 \left(\frac{M_2}{{\rm TeV}}\right)^2
\end{equation}
For the Chargino, the mass limit is even higher:
\begin{equation}
m_{gr}\leq 10^3{\rm GeV}
\end{equation}
\begin{equation}
m_{\tilde{W}^+}\geq 40 {\rm TeV}
\end{equation}
All of these regions are excluded by the requirement of gauge
coupling unification.

\subsection{Photino NLSP}

As we saw in the former section, hadronic constraints pushed the
mass of the NLSP different from a Bino one above the 10 TeV scale,
with the resulting conflict with gauge coupling unification.
Anyway, just looking at the electromagnetic constraints in
fig.\ref{Higgsino} , one can see that a particle that will not
decay hadronically at the time of Nucleosynthesis will be allowed
to be one order of magnitude lighter. This makes the photino,
$\tilde{A}=\cos(\theta_W)\tilde{B}+\sin(\theta_W)\tilde{W}_3$, a
natural
candidate to be a NLSP with Gravitino LSP.\\

Computing the relic abundance of a Photino is rather complicated,
even in Split Susy. The reason is that co-hannilation channels
with the charged Winos makes a lot of diagrams allowed. In order
to estimate the Photino relic abundance, we then observe that, because
of the fact that the Bino is very weakly interacting in Split
Susy, Photino annihilation channels will be dominated by the
contribution of the channels allowed by the Wino component. We
then quite reliably estimate the Photino relic abundance starting from the
formula for the relic abundance of a pure Wino particle we found before:
\begin{equation}
\Omega_{Wino NLSP}h^2=0.02\left(\frac{M_2}{{\rm TeV}}\right)^2
\end{equation}
and consider that the Photino has a Wino component equal to $\sin(\theta_W)$
So, for the Photino case we will have:
\begin{equation}
\Omega_{Photino
NLSP}\cong\frac{0.02}{\sin(\theta_W)^4}\left(\frac{M_{\tilde{A}}}{{\rm
TeV}}\right)^2\cong0.37 \left(\frac{M_{\tilde{A}}}{{\rm TeV}}\right)^2
\end{equation}

We then obtain the allowed region shown in fig.(\ref{Photino}).
The graph is very similar to the Higgsino case, with the
difference that the Electromagnetic constraints are less stringent
than the Hadronic ones. This allows to have the following
region:
\begin{equation}
700\ {\rm GeV}\leq M_{\tilde{A}}\leq 5\ {\rm TeV}
\end{equation}
the lightest part of which might be reachable at LHC. However, already if
we allow an hadronic branching ratio of $10^{-3}$, we see that the Photino NLSP
becomes excluded. So, we conclude that a Photino NLSP is in principle allowed,
but only if we fine tune it to be extremily close to a pure state of Photino.

\begin{figure}[!h]
\begin{center}
\includegraphics[scale=0.88]{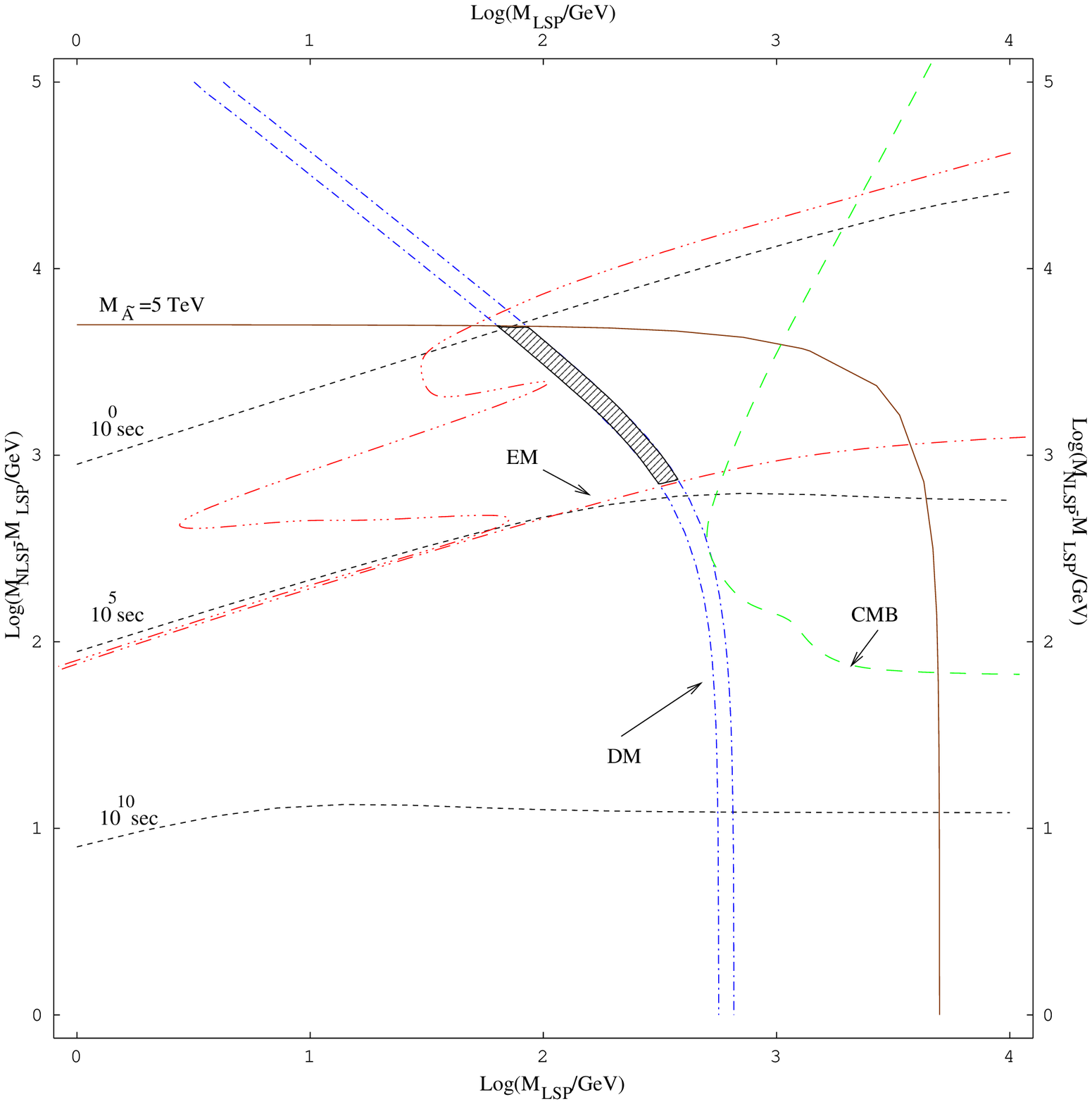}
\caption{Shaded is the allowed region for the Photino NLSP, Gravitino LSP.
The long dashed contour delimitates from the left the region excluded by CMB, the dash-dot-dot 
 contour
delimitates from above the region excluded by
the EM constraints from BBN in the case $B_h\sim0$, 
the dashed-dot-dot-dot
countor represents the same for $B_h\sim 10^{-3}$. The region within the 
dash-dot lines
represents the region where $\Omega_{DM}$ is within the
experimental limits; finally, we show the short dashed contours which represent where
the Photino decays to Gravitino
at 1 sec, $10^5$ sec, and $10^{10}$ sec, and the solid contour where the
Photino is 5 TeV heavy, which represents the upper limit for Gauge Couplig
Unification. We see that already for $B_h\sim 10^{-3}$ a Photino NLSP is 
excluded.
}\label{Photino}
\end{center}
\end{figure}

\subsection{Bino NLSP}

Bino NLSP is a very good candidate for avoiding all the cosmological 
constraints. In Split Susy, a
Bino NLSP is almost not interacting. For a pure Bino, the only interaction
which determines its relic abundance is the annihilation to Higgs
bosons through the exchange of an Higgsino. Since this cross
section is naturally very small, by eq.(\ref{Omega_NLSP}),
$\Omega_{NLSP}$ is very big. This means that, in order to create the
right amount of DM (see eq.(\ref{Omega_LSP})), we need to make the
Gravitino very light. And this is exactly what we need to do in
order to avoid the nucleosynthesis bounds. We conclude then that,
of the two directions to solve the DM and the nucleosynthesis
problems that we outlined in the former sections, a Bino NLSP would
naturally pick up the one which is the most favorable for LHC
detection. However, as we saw in the section on Gauge Coupling Unification
(see fig.\ref{Gut_Unification_Higgsino_fig}),
the Higgsino can not be much heavier than the Bino. This implies that 
a Bino like NLSP will have to have some Higgsino component unless it is 
very heavy and the off diagonal terms in the mass matrix are uniportant.
As a consequence, new annihilation channels opens up for the Bino NLSP through
its Higgsino component. This has the effect of diminishing the relic abundance
of a Bino NLSP with respect to the naive thinking we would have done 
if we neglected
the mixing. As a result, the cosmological constraints begin to play an 
important role in the region of the spectrum we are interested in, 
and, at the end, considering the upper limit from gauge coupling 
unification, exclude a Bino NLSP.\\

In order to compute the Bino NLSP relic abundance, we again modify the Dark 
Susy code\cite{Gondolo:2002tz}. The results are shown in fig.\ref{Bino_A}. As anticipated, we 
see that the relic 
abundance strongly depends on the ratio beteween the Bino and the Higgsino 
masses $M_1,\mu$. The relic abundance of the Bino NLSP becomes large enough
to avoid the cosmological constaints only for so large values of the ratio
between $\mu$ and $M_1$ which are not allowed by Gauge Coupling Unification.
We conclude, then, that a Bino NLSP is not allowed.

\begin{figure}[!h]
\begin{center}
\includegraphics[scale=1.0]{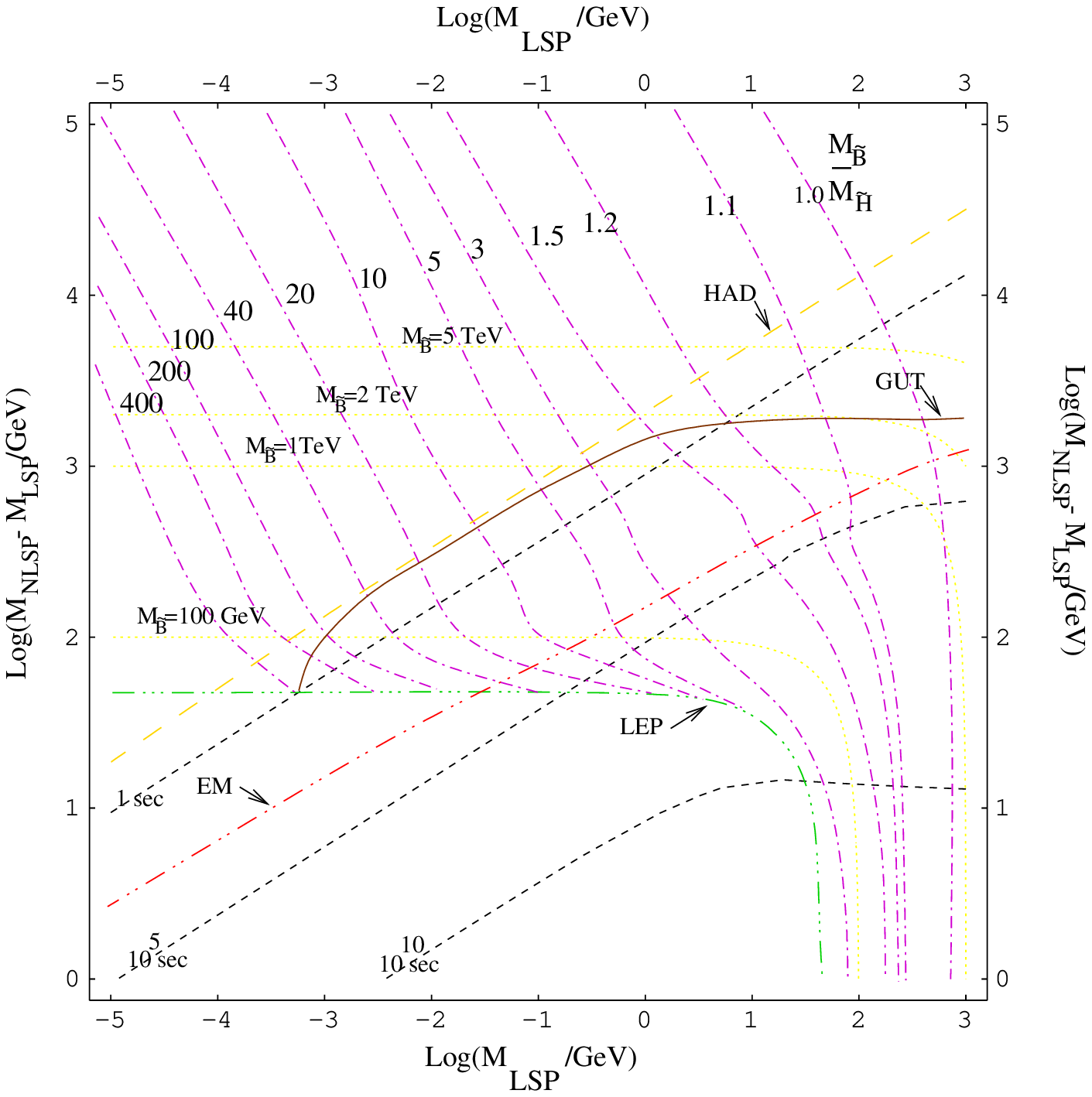}
\caption{Constraints for Bino NLSP. 
The long dashed countour delimitates from the left the
excluded region by the hadronic constraint from BBN,
the  dash-dot-dot contour represents the same for EM constraints from
BBN,
the dash-dot lines represents the ratio between the
Higgsino mass and the Bino mass necessary for $\Omega_{LSP}$ to be
equal to observed DM amount; the solid line represents the upper limit from
GUT, while the dash-dot-dot-dot line represents the lower limit from LEP;
we also show in short dashed the contours
where the Bino decays to Gravitino at 1 sec, $10^5$ sec, and
$10^{10}$ sec, and in dotted some characterstic contours for
the Bino mass.
CMB constraint plays no role here. We take $M_2\simeq 2 M_1$, as inferred from 
gaugino mass unification at the GUT scale \cite{Giudice:2004tc}, 
and we see that no allowed region 
is present. 
}\label{Bino_A}
\end{center}
\end{figure}

\section{Extradimansional LSP}
When we consider generic possibilities to break Susy, we can have,
further than the Gravitino, other fermions in a hidden sector
which are kept light
 by an R symmetry. The implications of these fermions as being the LSP
 can be quite different with respect to the case of Gravitino LSP, as we will
 see in this section.
 Here, we concentrate os Susy breaking in Extra Doimensions, where, as it was
 shown in \cite{Arkani-Hamed:2004fb}, it is very generic to
 expect a light fermion in the hidden sector.\\

In Susy breaking in Extra Dimension, one can break Susy with a
radion field, which gets a VEV. Its fermionic component, the
Goldstino, is then eaten by the Gravitino which becomes massive.
Even though at tree level there is no potential, one sees that at
one loop the Casimir Energy makes the radius instable. One can
compensate for this introducing some Bulk Hypermultiplets, finding
that, however, the cosmological constant is negative. Then, in
order to cancel this, one finds that he has to introduce another
source of symmetry breaking, a chiral superfield $X$ localized on
the brane (see for example \cite{Arkani-Hamed:2004fb}). This
represents a sort of minimal set to break Susy in Extra
Dimensions. If one protects the $X$ field interactions with a U(1)
charge, than one finds that the interactions with the SM particles
are all suppressed by the 5 dimensional Plank Mass, of the form:
\begin{equation}
\int d^4 \theta \frac{1}{M^2_5} X^\dag X Q^\dag Q
\end{equation}
This induces the following mass spectrum \cite{Arkani-Hamed:2004fb}:
\begin{equation}
m_{gr}\sim \frac{\pi M^3_5}{M^2_4}; m_S\sim \frac{\pi
M^5_5}{M^4_4}; m_i,\mu,m_{\psi_X}\sim\frac{\pi M^9_5}{M^8_4}
\end{equation}
where $M^2_4\sim r M^3_5$ are the 4 and 5 dimensional Plank
constants, and where $M_i$ are the gaugino masses.\\
It is quite natural to use the extradimension to lower the higher dimensional 
Plank mass
to the GUT scale, a la Horawa-Witten \cite{Horava:1995qa}, $M_5\sim M_{GUT}\sim
3\times10^{16}$ GeV. We have this range of scales\cite{Arkani-Hamed:2004fb}:
\begin{equation}
m_{gr}\sim 10^{13}{\rm GeV}; m_S\sim 10^9 {\rm GeV}; m_{{\rm
radion}}\sim 10^7 {\rm GeV}; M,\mu,m_{\psi_X}\sim 100 {\rm GeV}
\end{equation}
We notice that we have just reached the typical spectrum of Split
Susy, in a very natural way: we break Susy in Extra Dimension,
stabilize the moduli, and we introduce a further Susy breaking
source to compensate for the cosmological constant. We further notice that 
there is no much room to move the higher dimensional Plank mass $M_5$ away 
from the Horawa-Witten value. In fact, the fermion mass scales as 
$\left(\frac{M_4}{M_5}\right)^8$, so, a slight change of $M_5$ makes the
fermions of Split Susy generically either too heavy, 
making them excluded by gauge coupling unification, or too light, making
them conflict with collider bounds.\\

Concerning the
study of the LSP, we notice that the fermionic component of the
$X$ field we have to introduce in order to cancel the cosmological
constant is naturally light, of the order of the mass of the
Gauginos. So, it is worth to investigate the case in which this
fermion is the LSP, and how this case differs from the case in
which the LSP
is the Gravitino.\\

Concerning the DM abundance, nothing changes with respect to the
case of the gravitino LSP, so, we can keep the former results.\\

Next step it is to evaluate the decay time, to check if the
nucleosynthesis and CMB constraints play a role.\\

To be concrete, let us concentrate on the Higgsino NLSP. When the
Higgsino is heavier than the Higgs, the leading contribution to
the decay of the Higgsino will come from the tree level diagram
mediated by the operator\cite{Arkani-Hamed:2004fb}:
\begin{equation}
\int d^2\theta \frac{m^2X}{M^2_5}H_uH_d
\end{equation}
The decay time is then given by:
\begin{equation}
\tau=\frac{128}{\pi}\left(\frac{M_4}{\pi
M_5}\right)^8\left(1+\frac{m^2_{\psi_X}}{m^2_{\tilde{H}}}-\frac{m^2_h}
{m^2_{\tilde{H}}}\right)^{-1}
\left(\frac{(m^2_{\tilde{H}}+m^2_{\psi_X}-m^2_h)^2}{4
m^2_{\tilde{H}}}-m^2_{\psi_X}\right)^{-1/2}
\end{equation}
In the limit of $m_{\tilde{H}}\gg m_{\psi_X},m_h$ this expression
simplifies to:
\begin{equation}
\tau\sim \frac{128}{\pi^9 m_{\tilde{H}}}\left(\frac{M_4}{M_5}\right)^8
\end{equation}
Estimating with the number we just used before, we get:
\begin{equation}
\tau\sim 10^{-14} \left(\frac{{\rm TeV}}{m_{\tilde{H}}}\right) {\rm sec}
\end{equation}
This time is so long before nucleosynthesis, that all the
BBN constraints we found in the former case for the
Gravitino now disappear. Clearly, this statment is not affected if
we vary $M_5$ in the very small window allowed by the restrictions on the
fermionic superpartners' spectrum. Basically, in this mass regime, the only
constraint which will apply will be the one from $\Omega_{DM}$. As
we can see form the formula for the higgsino relic abundance, 
the region where Higgsino is lighter that Higgs is not
relevant, and is excluded by the constraint on Dark Matter abundance.\\

So, we conclude that nucleosynthesis and CMB constraints do not
apply in the case in which the LSP is the field $\psi_X$, and the Higgsino 
is the NLSP, the only
constraints which applies are the one coming from $\Omega_{DM}$ and the one
coming from gauge coupling unification. In
fig.\ref{Higgsino_extra}, we show the allowed region. While the
full region is quite large, and covers a rather large phase space,
there is a region where the Higgsino is rather light, $\sim
2$ TeV, and the mass of the field $\psi_X$ is
constrained quite precisely to be around $2$ TeV. The
region is bounded from above by the limit on gauge coupling
unification at around 18 TeV, as in this case we must allow also for
anomaly mediated initial conditions for Gauginos mass at the
intermediate scale. This region is not extremely attractive for
LHC.\\

\begin{figure}[!h]
\begin{center}
\includegraphics[scale=1.1]{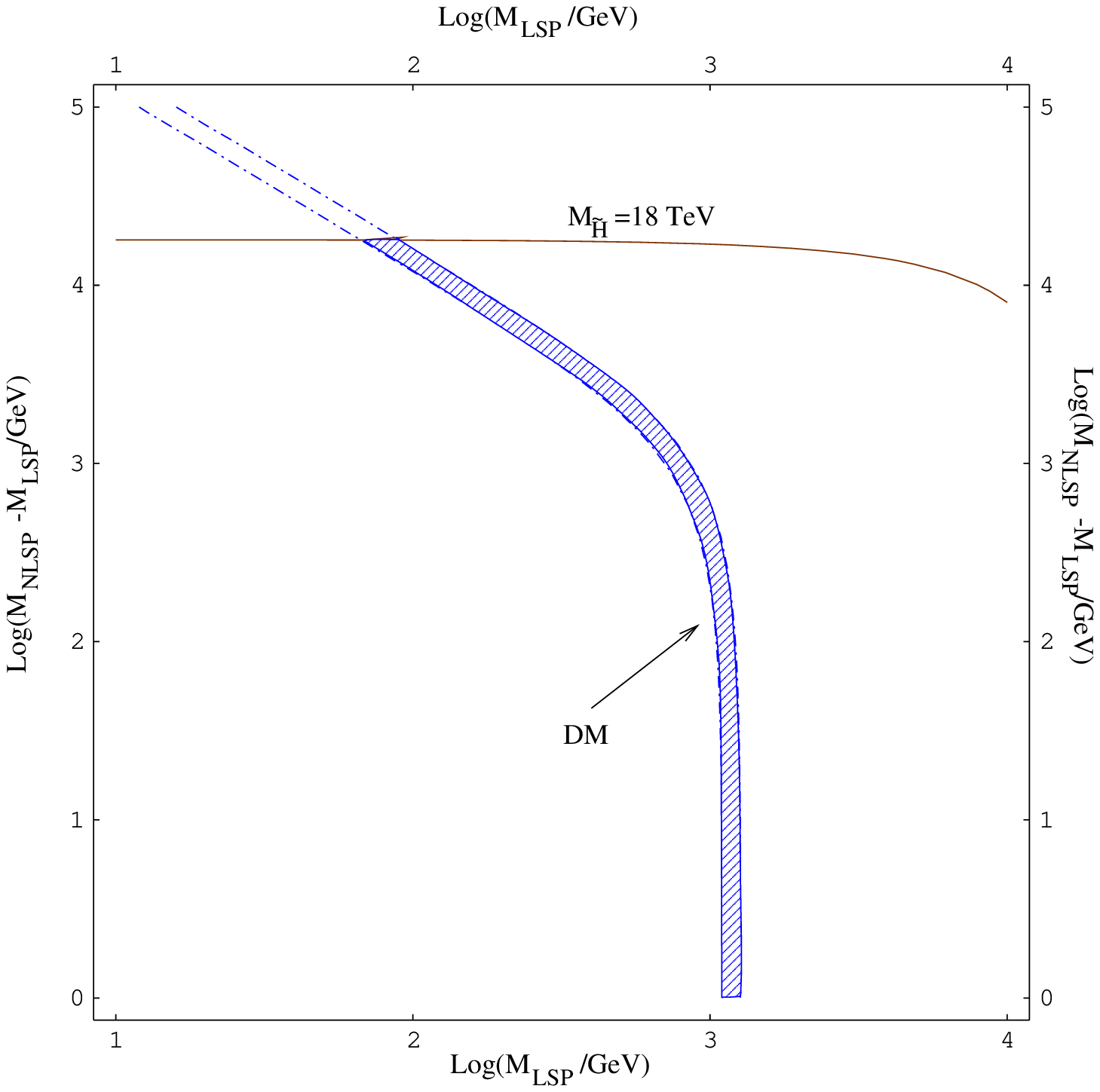}
\caption{Shaded is the allowed region for the Neutral Higgsino NLSP, $\psi_X$
LSP. Since there are no constraints from CMB and BBN,
the only constraints come from $\Omega_{DM}$, which delimitates
the region within the dash-dot lines, and Gauge Coupling
Unification, which set the upper bound of 18 TeV with the solid line. 
For Neutral Wino and
Chargino NLSP the result is very similar.}
\label{Higgsino_extra}
\end{center}
\end{figure}

For gaugino NLSP, 
the situation is very similar, as the decay of the gauginos to the field 
$\psi_X$ is mediated by the same kind of operator as for the Higgisino case
\cite{Arkani-Hamed:2004fb}:
\begin{equation}
\int d^2\theta \frac{m^2X}{M^2_5}WW
\end{equation}
where $W$ is the gaugino vector supermultiplet. Clearly, again in this case,
the decay time will be way before the time of BBN. 
In this cases, the curves that delimitate the
allowed region are practically identical to the one of the
Higgsino, with the only difference that the region where the NLSP
is the lightest and it is practically degenerate with the $\psi_X$,
correspont to an higher mass of $\sim 3$ TeV, more difficult to see
at LHC.\\

Similarly occurs for the Bino NLSP, with the only difference that 
the region allowed by the Dark Matter constraint is a bit different with
respect to the case of Higgsino and Wino. We obtain the allowed region shown in
fig.\ref{Bino_extra_A}, where we see
that the spectrum is very light, with Bino and Higgsino starting at tens of GeV
, and gluinos at 200 GeV, with $\psi_X$ in the
range $10^{1}-10^3$ GeV . This is a very good region for LHC. 
Notice that the upper limit on the Bino mass is again 5 TeV,
as in the case the Bino is the NLSP,  we can not have anomaly mediated
initial conditions
for Gaugino mass at the intermediate scale.\\

\begin{figure}[!h]
\begin{center}
\includegraphics[scale=1.15]{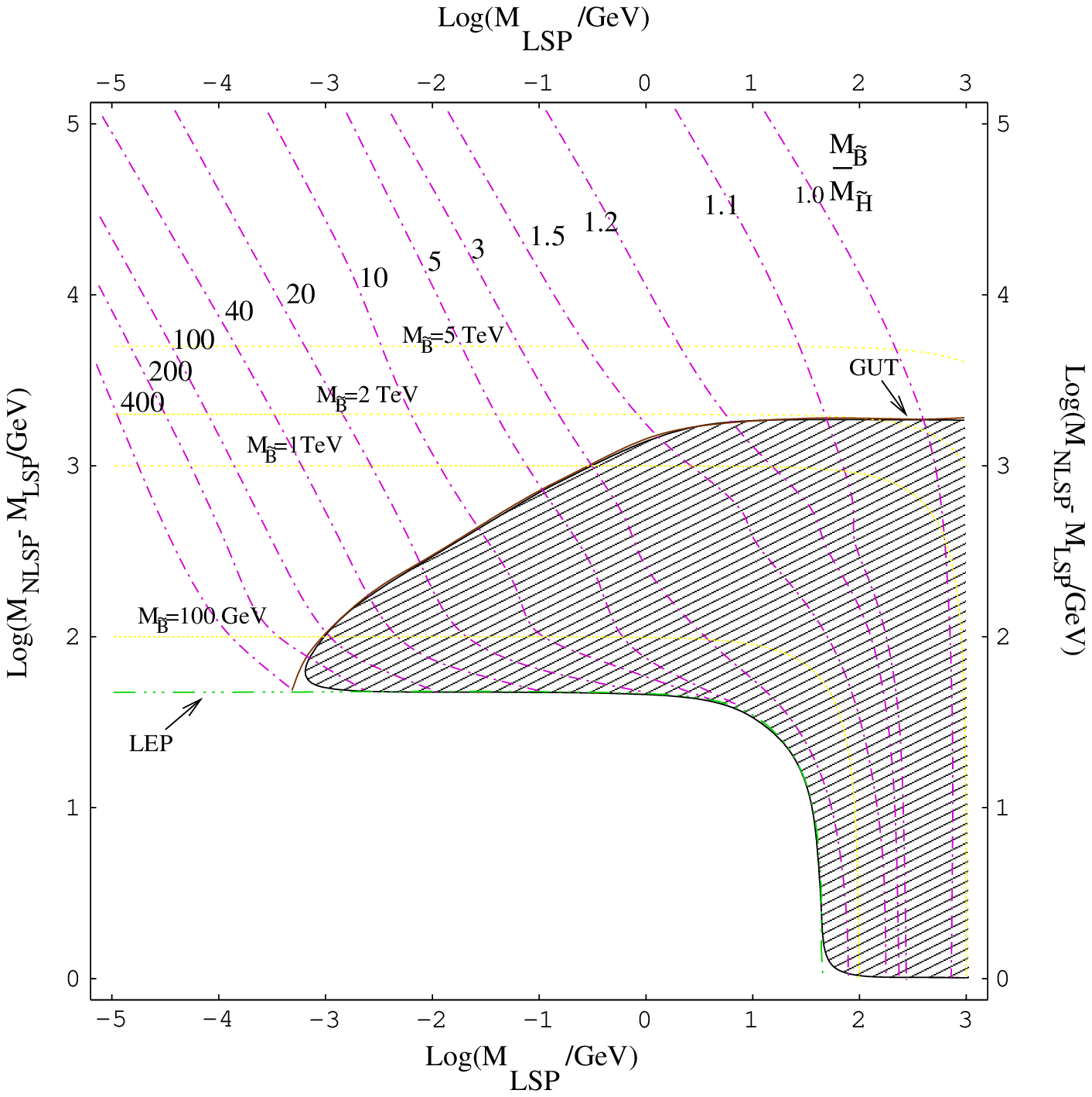}
\caption{Shaded is the allowed region for the Bino NLSP, $\psi_X$ LSP.
The
dash-dot lines represent the ratio between the Higgsino mass and
the Bino mass in order for $\Omega_{LSP}$ to be equal the observed
amount of DM. The dotted lines are some characteristic
countors for the Bino mass. The solid line is the upper limit from GUT,
while the dash-dot-dot-dot line is the lower limit from LEP. We take $M_2\simeq 2 M_1$.}
\label{Bino_extra_A}
\end{center}
\end{figure}

\section{Conclusions}
In Split Susy, the only two motivations to expect new physics at
the TeV scale are given by the requirement that gauge coupling
unification is achieved, and, mostly, that the stable LSP makes up
the Dark Matter of the Universe. This is true in the standard
scenario where the LSP is a neutralino. Here we have investigated
the other main alternative for the LSP, that is that the LSP is
constituted by a hidden sector particle. A natural candidate for
this is the Gravitino, which here we studied quite in detail.
Nevertheless, it is true that among the different possibilities we
have in order to break Susy, one can expect the appearence in the
spectrum of another light fermion protected by R symmetry. Here,
as an example, we study the case of a light fermion arising in
Susy breaking in Extra Dimension.
\\

The requirement to obtain gauge coupling unification limits the
masses for the fermions to be less than 5 TeV or 18 TeV, according
to the different initial conditions for the Gaugino masses at the
the intemediate scale $\tilde{m}$.\\

In this range of masses, we have seen how constraints from
Nucleosynthesis put strong limits on the allowed region. In fact,
there are two competing effects: in order to avoid Nuclosynthesis
constraints, the NLSP must decay to the LSP early, and this is
achieved creating a big hierarchy between the NLSP and the LSP. On
the other hand, this hierarchy tends to diminish the produced
$\Omega_{LSP}$, and in order to compensate for it, the NLSP tends
to be heavy. This goes against the constraints from gauge coupling
unification. This explains why
a large fraction of the parameter space is excluded.\\

The details depend on the particular LSP and NLSP.

\subsection{Gravitino LSP}

Gravitino LSP forces us to consider Gaugino Mass Unification at
the intermediate scale as initial condition. This implies that we
have to live with the more restrictive upper limit on fermionic
masses from gauge coupling unification: 5 TeV.\\

At the same time, the typical decay time of an NLSP to the
Gravitino is at around 1 sec, and this goes exactly into the
region where constraints from Nucleosynthesis on ElectroMagnetic
and Hadronic decays apply.\\

The final result is that only if the NLSP is very pure Photino like, then
the Gravitino can be the LSP, with a Photino between 700 GeV and 5 TeV. 
If the NLSP is different by this case, then the 
Gravitino {\it can not} be the LSP. The reason is that a very Photino like 
NLSP can avoid the constraints on BBN on hadronic decays, 
which are much more stringent
than the ones coming from electromagnetic decays, and so it can be light
enough to avoid the upper limit on its mass 
coming from gauge coupling unification.\\

This is very good news for the detactability of Split Susy at LHC.
In fact, if the Gravitino was the LSP, than the NLSP could have been 
much heavier than around 1 TeV, making detection very difficult. In this paper
we show that this possibility is almost excluded.\\

\subsection{ExtraDimensional LSP}
Following the general consideration that in breaking Susy we might
expect to have some fermion other than the Gravitino in the hidden
sector which is kept light by an R symmetry, we have studied also
the possibility that the fermionic component of a chiral field
which naturally arises in Extra Dimensional Susy breaking is the
LSP. In this case the time of decay is so early that no
Nucleosynthesis bounds apply, so, the only constraints applying
are those from
Dark Matter and gauge coupling unification.\\

Concerning gauge coupling unification, in this case we must
consider the possibility that the Gaugino Mass initial conditions
are also those from anomaly mediation. This implies that we have
to give the upper limit of 18 TeV to Fermions' mass.\\

Having said this, the lower bound on the mass is given by the DM
constraint. In fact, it is clear that $\Omega_{NLSP}$ must be
greater than $\Omega_{LSP}$. As a consequence, the Mass of the
NLSP has to be greater than the one found in
\cite{Giudice:2004tc,Pierce:2004mk} for the case in which these
particles where the LSP. As a consequence, Charginos
and Neutralinos NLSP are in general allowed, but they
are restricted to be heavier than 1 TeV. It is quite interesting that,
in this cases, the LSP is restricted to be in the range
100-1000 GeV.\\

Again, there is an exception: the Bino.
In this case, the LSP and NLSP are  much lighter than in
the case of the others Neutralinos NLSP, with an NLSP as light as
a few decades of GeV.

\section*{Acknowledgments}
I would like to thank  Nima Arkani-Hamed and Paolo Creminelli, who
inspired me the problem, and without whose constant help I would
have not been able to perform this work. I would like to thank also
Alberto Nicolis for interesting conversations, and Rakhi Mahbubani
and Aaron Pierce for their help with Dark Susy.

\end{document}